\documentclass[11pt]{article}
\newcommand{\be}{\begin{equation}}
\newcommand{\ee}{\end{equation}}
\newcommand{\bea}{\begin{eqnarray}}
\newcommand{\eea}{\end{eqnarray}}

\begin{document}
\vspace{.5in} 
\begin{center} 
{\LARGE{\bf Integrable and Superintegrable Klein-Gordon and Schr\"odinger 
Type Dimers}}
\end{center} 

\vspace{.3in}
\begin{center} 
{\LARGE{\bf Avinash Khare}} \\ 
{Physics Department, Savitribai Phule Pune University \\
Pune, India 411007}
\end{center} 

\begin{center} 
{\LARGE{\bf Avadh Saxena}} \\ 
{Theoretical Division and Center for Nonlinear Studies, Los
Alamos National Laboratory, Los Alamos, NM 87545, USA}
\end{center} 

\vspace{.9in}
{\bf {Abstract:}}  

A $PT$-symmetric dimer is a two-site nonlinear oscillator or a nonlinear 
Schr\"odinger dimer where one site loses and the other site gains energy 
at the same rate.  We present a wide class of integrable oscillator type 
dimers whose Hamiltonian is of arbitrary even order. Further, we also 
present a wide class of integrable and superintegrable nonlinear Schr\"odinger 
type dimers  where again the Hamiltonian is of arbitrary even order.  

\newpage 
  
\section{Introduction} 

In recent years, parity-time reversal or $PT$-symmetry \cite{ben} has received 
widespread attention from both the physics and mathematics community. 
While the original application of $PT$-symmetry was in quantum
mechanics, soon researchers realized \cite{opt} that optics can provide 
a fertile ground to test some of these ideas experimentally \cite{opex}. It was 
realized there that ubiquitous loss can be countered by an 
overwhelming gain in order to create a $PT$-symmetric setup. For example,
such a setup in wave guide dimers paved the way for numerous developments
especially in the context of nonlinear systems. Not surprisingly, most of
the activity is centered around nonlinear Schr\"odinger type systems. This
is understandable since Schr\"odinger model is on a similar footing as
paraxial approximation in the Maxwell equations. However, recently,
motivated by experimental activity \cite{oscdim}, for example, in the areas  
of electronic circuits \cite{ele}, whispering gallery modes, micro-cavities 
\cite{cav}, etc., significant theoretical progress has been made. Some recent 
theoretical activity has also centered on the oscillator type dimers \cite{kha}. 

In a recent publication, Barashenkov et al. \cite{ba} have written down 
integrable Schr\"odinger type dimer models with a quartic Hamiltonian. The
importance of such integrable dimers cannot be overemphasized. They can
act as prototype systems (analogous to integrable models in classical 
mechanics). Their work raises several questions. For example, are there
oscillator-type integrable dimer models? And if they exist, then within the
rotating wave approximation (RWA), do they lead to integrable Schr\"odinger
type dimer models? Can one generalize and obtain the integrable oscillator
as well as Schr\"odinger type dimers in case the Hamiltonian is of arbitrary 
even order? And finally, are there superintegrable dimer models? By 
superintegrable we mean that the system has $2n-1$ constants of motion 
in involution where $n$ is the degrees of freedom of the classical system (so 
that the phase space is $2n$ dimensional) \cite{super}.  Note that for an 
integrable system there are $n$ constants of motion in involution. 

The purpose of this paper is to answer many of the questions raised here.
In particular, we present a class of integrable oscillator type dimer
models where the Hamiltonian is of arbitrary even order. We further show 
that within the RWA, these integrable dimers lead to superintegrable 
Schr\"odinger type dimers. We generalize and write down the most general 
superintegrable Schr\"odinger type dimers where the Hamiltonian is of 
arbitrary even order. Finally, we also present a class of integrable 
(but not superintegrable) Schr\"odinger type dimers where the
Hamiltonian is of arbitrary even order.

\section{Oscillator Type Integrable Dimers}

To motivate the discussion, 
let us start from the following coupled equations   
\be\label{1}
\ddot{u} = - u + kv + \gamma \dot{u} +\epsilon v^3 + \delta vu^2 ,
\ee
\be\label{2}
\ddot{v} = - v +ku  - \gamma \dot{v} + \epsilon u^3 + \delta u v^2 .
\ee

We now show that in the special case when $k = \epsilon =0$, it is an integrable
system. It is easy to see that these coupled equations can be derived from the 
Hamiltonian
\be\label{3}
H = p_{u} p_{v} +\frac{\gamma}{2} (u p_{u}-vp_{v})+(1-\frac{\gamma^2}{4})u v
-\frac{k}{2}(u^2+v^2) -\frac{\epsilon}{4} (u^4 + v^4) 
-\frac{\delta}{2} u^2 v^2\,.
\ee
In this case the momenta are given by: $p_{u} = \dot{v}+\gamma v/2,~p_{v} = 
\dot{u}-\gamma u/2$. It is remarkable that for arbitrary $k, \epsilon$ and $\delta$, 
this is a Hamiltonian system. 

Remarkably, when $k=\epsilon =0$ but arbitrary $\delta$, there is another 
constant of motion 
\be\label{4}
C= u\dot{v}-v\dot{u}+\gamma uv\,.
\ee

Thus coupled Eqs. (\ref{1}) and (\ref{2}) with $k = \epsilon =0$ is an 
integrable PT-invariant oscillator type dimer system with two constants of 
motion $H$ and $C$. As far as we are aware of, this is the first known 
PT-invariant, integrable nonlinear oscillator dimer system.
This discussion can be immediately generalized and one can obtain a wide
class of PT-invariant integrable Klein-Gordon (KG) dimer systems.

In particular, let us consider the following coupled PT-invariant 
KG dimer system 
\be\label{5}
\ddot{u} = - u + kv + \gamma \dot{u} +\epsilon v^{2n+1} 
+ \delta v^n u^{n+1}\,,~~n= 1,2,3,...\,,
\ee
\be\label{6}
\ddot{v} = - v +ku  - \gamma \dot{v} + \epsilon u^{2n+1} 
+ \delta u^n v^{n+1}\, . 
\ee
It is easy to see that these coupled equations can be derived from the 
Hamiltonian
\be\label{7}
H = p_{u} p_{v} +\frac{\gamma}{2} (u p_{u}-vp_{v})+(1-\frac{\gamma^2}{4})u v
-\frac{k}{2}(u^2+v^2) -\frac{\epsilon}{2n+2} (u^{2n+2} + v^{2n+2}) 
-\frac{\delta}{n+1} u^{n+1} v^{n+1}\,.
\ee
In this case, $p_{u} = \dot{v}+\gamma v/2,~p_{v} = \dot{u}-\gamma u/2$. 
It is remarkable that for arbitrary $k, \epsilon$ and $\delta$, this is a 
Hamiltonian system. 
It is easy to check that in the case $\epsilon= k =0$ but $\delta$ arbitrary, 
there is another constant of motion given by Eq. (\ref{4}).

Thus coupled Eqs. (\ref{5}) and (\ref{6}) with $k = \epsilon =0$ but arbitrary
$\delta$ and $n$ is an 
integrable PT-invariant KG dimer system with two constants of motion 
$H$ and $C$ for arbitrary positive integer values of $n$.  

One can even further generalize and have an even wider class of PT-invariant
integrable oscillator type dimer systems. In particular, consider the following 
coupled equations
\bea\label{8}
&&\ddot{u} = - u + kv + \gamma \dot{u} +\epsilon_{1}  v^{2n_{1}+1}
+\epsilon_{2} v^{2n_{2}+1} \nonumber \\
&&+\delta_{1} v^{n_1} u^{n_{1}+1}
+\delta_{2} v^{n_2} u^{n_{2}+1}\,,
\eea
\bea\label{9}
&&\ddot{v} = - v + ku - \gamma \dot{v} +\epsilon_{1}  u^{2n_{1}+1}
+\epsilon_{2} u^{2n_{2}+1} \nonumber \\
&&+\delta_{1} u^{n_1} v^{n_{1}+1}
+\delta_{2} u^{n_2} v^{n_{2}+1}\, .
\eea
 These coupled equations can be derived from the Hamiltonian
\bea\label{10}
&&H_2 = p_{u} p_{v} +\frac{\gamma}{2} (u p_{u}-vp_{v})
+(1-\frac{\gamma^2}{4})u v -\frac{k}{2}(u^2+v^2) 
-\frac{\epsilon_1}{2n_1 +2} (u^{2n_{1}+2} + v^{2n_{1}+2}) \nonumber \\
&&-\frac{\epsilon_2}{2n_2 +2} (u^{2n_{2}+2} + v^{2n_{2}+2}) 
-\frac{\delta_1}{n_1 +1} u^{n_{1}+1} v^{n_{1}+1}
-\frac{\delta_2}{n_2 +1} u^{n_{2}+1} v^{n_{2}+1} . 
\eea
In this case, $p_{u} = \dot{v}+\gamma v/2,~p_{v} = \dot{u}-\gamma u/2$. 
It is easy to check that in case $\epsilon_{1,2} = k =0$ but 
$\delta_{1,2}$ arbitrary, then $C$ as defined by Eq. (\ref{4})
continues to be another constant of motion.

Thus coupled Eqs. (\ref{8}) and (\ref{9}) with $k = \epsilon_{1,2} =0$ is an 
integrable PT-invariant KG dimer system with two constants of motion 
$H_{2}$ and $C$ for arbitrary positive integer values of $n_1, n_2$
and for arbitrary values of $\delta_{1,2}$.
Generalization to sum of several terms of the form $\epsilon_i v^{n_i}$
and $\delta_i v^{(n_i -1)/2} u^{(n_i +1)/2}$ in $\ddot{u}$ and 
appropriate terms in $\ddot{v}$ is straightforward.

Summarizing, we have obtained a wide class of PT-invariant integrable
oscillator type dimer systems. It would thus be
worthwhile to study these coupled models in detail. We hope to do that in the
near future. 

\section{Superintegrable Schr\"odinger Type Dimers}

We now show that unlike the oscillator type dimers, one can in fact
construct a $n+1$ parameter family of superintegrable Schr\"odinger type dimers 
corresponding to an arbitrary even order Hamiltonian $H$. 
 
To motivate the discussion, we first write down the 4th order
Hamiltonian and show that it is superintegrable and then generalize the
discussion to arbitrary even order. 

Consider the following quartic cross-gradient Hamiltonian 
(which gives rise to cubic dimer equations) 
\bea\label{x1}
&&H = A(\phi_{1}^{*} \phi_2 + \phi_{2}^{*} \phi_1)  
+ i\Gamma(\phi_{2}^{*} \phi_1 - \phi_{1}^{*} \phi_2)  
\nonumber \\
&&+D_1(\phi_{1}^{*} \phi_2 +\phi_{2}^{*} \phi_1)^2 +D_2 |\phi_1|^2 |\phi_2|^2\,.
\eea
As noted, this gives rise to the cubic dimer equations
\be\label{x2}
i\frac{d\phi_1}{dt} = A \phi_1 + i\gamma \phi_1 
+2D_1 (\phi_{1}^{*} \phi_2 + \phi_{2}^{*} \phi_1) \phi_1 
+D_2 |\phi_1|^2 \phi_2\,,
\ee 
\be\label{x3}
i\frac{d\phi_2}{dt} = A \phi_2 - i\gamma \phi_2 
+2D_1 (\phi_{1}^{*} \phi_2 + \phi_{2}^{*} \phi_1) \phi_2 
+D_2 |\phi_2|^2 \phi_1\, .
\ee 

Notice that in this case the canonical coordinate-momentum pairs are 
$\phi_1, \dot{\phi}_{2}^{*}$ and $\phi_2, \dot{\phi}_{1}^{*}$ so that
in this case the Hamilton's equations are
\be\label{x4}
i\frac{d\phi_1}{dt} = \frac{\partial H}{\partial \phi_{2}^{*}}\,,~~~
i\frac{d\phi_2}{dt} = \frac{\partial H}{\partial \phi_{1}^{*}}\,.
\ee
It may be noted here that in order to distinguish the oscillator type
and Schr\"odinger type dimers, we use different symbols $\phi_1, \phi_2$
instead of $u,v$ which we used for the oscillator type dimers.

In order to study the question of integrability of the Schr\"odinger type dimers,
we introduce the Stokes variables
\be\label{xa}
X= \phi_{1}^{*} \phi_2 + \phi_{2}^{*} \phi_1\,,~~Y = i(\phi_{1}^{*} \phi_2 - 
\phi_{2}^{*} \phi_1)\,,~~ Z = |\phi_1|^2 - |\phi_2|^2\,,~~R = |\phi_1|^2 + 
|\phi_2|^2\,.
\ee 

Using the field equations, it is easy to check that the variables $X, Y$
as defined by Eq. (\ref{xa}) are time independent, i.e. $\dot{X} = 
\dot{Y} = 0$. Thus, this dimer system is superintegrable with the three
constants of motion being $H, X, Y$ as defined by Eqs. (\ref{x1}) and (\ref{xa}). 

Generalization to arbitrary even order is now straightforward. In 
particular, so long as the cross-gradient Hamiltonian only consists of 
the arbitrary combinations of the two invariants $X$ and 
$|\phi_1|^2 |\phi_2|^2$, then it will always be superintegrable with the
three constants of motion being $H, X, Y$. For example,  
the most general $4n$'th order cross-gradient superintegrable Hamiltonian 
has $n+1$ arbitrary parameters $D_i$ and is given by
\bea\label{x5}
&&H = A(\phi_{1}^{*} \phi_2 + \phi_{2}^{*} \phi_1)  
+ i\gamma(\phi_{2}^{*} \phi_1 - \phi_{1}^{*} \phi_2)  
\nonumber \\
&&+D_1 (\phi_{1}^{*} \phi_2 +\phi_{2}^{*} \phi_1)^{2n} 
+D_2 |\phi_1|^2 |\phi_2|^2 (\phi_{1}^{*} \phi_2 + \phi_{2}^{*} \phi_1)^{2n-2}
+...+D_{n+1} |\phi_1|^{2n} |\phi_2|^{2n}\,.
\eea

Similarly, the most general $(4n+2)$'th order cross-gradient superintegrable 
Hamiltonian too has $n+1$ arbitrary parameters $D_i$ and is given by
\bea\label{x6}
&&H = A(\phi_{1}^{*} \phi_2 + \phi_{2}^{*} \phi_1)  
+ i\gamma(\phi_{2}^{*} \phi_1 - \phi_{1}^{*} \phi_2)  
\nonumber \\
&&+D_1 (\phi_{1}^{*} \phi_2 +\phi_{2}^{*} \phi_1)^{2n+1} 
+D_2 |\phi_1|^2 |\phi_2|^2 (\phi_{1}^{*} \phi_2 + \phi_{2}^{*} \phi_1)^{2n-1}
+...+D_{n+1} |\phi_1|^{2n} |\phi_2|^{2n}(\phi_{1}^{*} \phi_2 
+\phi_{2}^{*} \phi_1)\,. 
\eea

For both these cases using Eqs. (\ref{x4}) it is straightforward to write equations 
of motion and show that $\dot{X} = \dot{Y} = 0$. 

It is worth pointing out that if one considers the integrable oscillator type 
dimer models discussed in the previous section, within the rotating wave approximation,
then they in fact lead to superintegrable Schr\"odinger type dimer models
as discussed here with the three constants of motion being cross-gradient
$H$, $X$ and $Y$.
 
\section{Wide Class of Schr\"odinger Type Integrable (but not Superintegrable) 
Dimer Models Based on Cross-Gradient Hamiltonians}

We have thus presented a $n+1$ parameter family of superintegrable, 
Schr\"odinger type dimers with even order cross-gradient Hamiltonians
of order $4n$ or $4n+2$. It is
then natural to ask if we can also obtain integrable (but not 
superintegrable) dimer models with arbitrary even order 
cross-gradient Hamiltonians. We now show that this is indeed possible 
and that we get either a $(n+1)^2$ or a $(n+1)(n+2)$ parameter family
of cross-gradient Hamiltonians of order $4n$ or $4n+2$, respectively 
with $n=1,2,3, ...$. Basically, one finds that the corresponding Hamiltonian
can be any combination of the three invariants $X, R$ and 
$|\phi_1|^2 |\phi_2|^2$.  For example, the 4th order cross-gradient integrable
Hamiltonian is given by
\bea\label{t1}
&&H = A(\phi_{1}^{*} \phi_2 + \phi_{2}^{*} \phi_1)  
+ i\gamma(\phi_{2}^{*} \phi_1 - \phi_{1}^{*} \phi_2)  
\nonumber \\
&&+D_1(\phi_{1}^{*} \phi_2 +\phi_{2}^{*} \phi_1)^2 +D_2 |\phi_1|^2 |\phi_2|^2
+D_3(\phi_{1}^{*} \phi_2 +\phi_{2}^{*} \phi_1)(|\phi_1|^2+ |\phi_2|^2)
+D_4 (|\phi_1|^2+|\phi_2|^2)^2\,. 
\eea
Using Eqs. (\ref{x4}) it is easy to write down the corresponding equations of motion 
and show that only $\dot{X} =0$ thereby obtaining an integrable
but not superintegrable cross-gradient Hamiltonian of 4th order. 

Generalization to any arbitrary cross-gradient Hamiltonian of order $4n$ 
is straightforward. It is easy to check that the corresponding 
$4n$'th order cross-gradient Hamiltonian is given by
\bea\label{t2}
&&H = A(\phi_{1}^{*} \phi_2 + \phi_{2}^{*} \phi_1)  
+ i\gamma(\phi_{2}^{*} \phi_1 - \phi_{1}^{*} \phi_2)  
\nonumber \\
&&+D_1(\phi_{1}^{*} \phi_2 +\phi_{2}^{*} \phi_1)^{2n} 
+D_2 (|\phi_1|^2 |\phi_2|)^{n}
+D_3(\phi_{1}^{*} \phi_2 +\phi_{2}^{*} \phi_1)^{2n-1}(|\phi_1|^2+ |\phi_2|^2)
\nonumber \\
&&+...+D_{(n+1)^2} (|\phi_1|^2+|\phi_2|^2)^{2n}\,,
\eea
having $(n+1)^2$ arbitrary parameters and for this case one can show
that only $\dot{X} =0$ thereby obtaining an integrable
but not superintegrable cross-gradient Hamiltonian of $4n$'th order with
$(n+1)^2$ parameters.

In the same way, it is easy to check that the ($4n+2$)'th order 
integrable Hamiltonian is given by
\bea\label{t3}
&&H = A(\phi_{1}^{*} \phi_2 + \phi_{2}^{*} \phi_1)  
+ i\gamma(\phi_{2}^{*} \phi_1 - \phi_{1}^{*} \phi_2)  
\nonumber \\
&&+D_1(\phi_{1}^{*} \phi_2 +\phi_{2}^{*} \phi_1)^{2n+1} 
+D_2 (|\phi_1|^2 |\phi_2|)^{n}(\phi_{1}^{*} \phi_2 +\phi_{2}^{*} \phi_1) 
+D_3(\phi_{1}^{*} \phi_2 +\phi_{2}^{*} \phi_1)^{2n}(|\phi_1|^2+ |\phi_2|^2)
\nonumber \\
&&+...+D_{(n+1)(n+2)} (|\phi_1|^2+|\phi_2|^2)^{2n+1}\,,
\eea
having $(n+1)(n+2)$ arbitrary parameters and for this case too 
only $\dot{X} =0$ thereby obtaining an integrable
but not superintegrable cross-gradient Hamiltonian of ($4n+2$)'th order.
 
\section{Integrable Schr\"odinger Type Dimers Based on Straight-Gradient
Hamiltonians}

So far we have obtained integrable as well as superintegrable Schr\"odinger
type dimer models of arbitrary even order. All these models are based on
the so called cross-gradient Hamiltonians with the canonical structure 
as given by Eqs. (\ref{x4}). It is then worth inquiring if there are
integrable as well as superintegrable Schr\"odinger type dimer models
which are based on direct-gradient Hamiltonians, i.e. in this case 
the canonical coordinate-momentum pairs would be 
$\phi_1, \dot{\phi}_{1}^{*}$ and $\phi_2, \dot{\phi}_{2}^{*}$ so that
in this case the Hamilton's equations would be
\be\label{y1}
i\frac{d\phi_1}{dt} = \frac{\partial H}{\partial \phi_{1}^{*}}\,,~~~
i\frac{d\phi_2}{dt} = \frac{\partial H}{\partial \phi_{2}^{*}}\,.
\ee

In this connection, it is worth pointing out that recently, 
Barashenkov et al. \cite{ba} have presented the most general
quartic integrable Schr\"odinger type dimer model based on the 
straight-gradient quartic 
Hamiltonian. In this model the two constants of motion are $H$ and $X$. 
We now generalize their results and write down the most general integrable
dimer models of arbitrary even order based on the straight-gradient 
Hamiltonians. 

As shown by Barashenkov et al. \cite{ba}, the most general quartic integrable 
straight-gradient Hamiltonian (which gives rise to cubic dimer equations) 
is given by
\bea\label{y2}
&&H = A(\phi_{1}^{*} \phi_2 + \phi_{2}^{*} \phi_1)  
+ B(|\phi_1|^2+ |\phi_2|^2) 
+ i\gamma(|\phi_1|^2 - |\phi_2)|^2)  
\nonumber \\
&&+D_1(\phi_{1}^{*} \phi_2 +\phi_{2}^{*} \phi_1)^2 
+D_2 (\phi_{1}^{*} \phi_2 +\phi_{2}^{*} \phi_1)
(|\phi_1|^2+|\phi_2|^2)+D_3(|\phi_1|^2+\phi_2|^2)^2\,.
\eea
Using Eqs. (\ref{y1}), this gives rise to the cubic dimer equations
\bea\label{y2}
&&i\frac{d\phi_1}{dt} = A \phi_1 + B \phi_2 + i\gamma \phi_1 
+2D_1 (\phi_{1}^{*} \phi_2 + \phi_{2}^{*} \phi_1) \phi_2 \nonumber \\
&&+D_3(|\phi_1|^2+|\phi_2|^2)\phi_2 +D_3 (\phi_{1}^{*} \phi_2 
+ \phi_{2}^{*} \phi_1) \phi_1+2D_4 (|\phi_1|^2+|\phi_2|^2) \phi_1\,, 
\eea 
\bea\label{y3}
&&i\frac{d\phi_2}{dt} = A \phi_2 + B \phi_1 - i\gamma \phi_2 
+2D_1 (\phi_{1}^{*} \phi_2 + \phi_{2}^{*} \phi_1) \phi_1 \nonumber \\
&&+D_3(|\phi_1|^2+|\phi_2|^2)\phi_1 +D_3 (\phi_{1}^{*} \phi_2 
+ \phi_{2}^{*} \phi_1) \phi_2+2D_4 (|\phi_1|^2+|\phi_2|^2) \phi_2\,. 
\eea 
 Using these field equations, it is easy to check that the variable $X$
as defined by Eq. (\ref{xa}) is time independent, i.e. $\dot{X} = 0$
and hence it is an integrable system with a straight-gradient Hamiltonian.
 
Generalization to an arbitrary even order is now straightforward. In 
particular, the most general $2n$'th order straight-gradient integrable 
Hamiltonian is given by
\bea\label{y7}
&&H = A(\phi_{1}^{*} \phi_2 + \phi_{2}^{*} \phi_1)  
+ B(|\phi_1|^2+ |\phi_2|^2) 
+ i\gamma(|\phi_1|^2 - |\phi_2)|^2)  
\nonumber \\
&&+D_1 (\phi_{1}^{*} \phi_2 +\phi_{2}^{*} \phi_1)^{n} 
+D_2 (|\phi_1|^2+|\phi_2|^2) (\phi_{1}^{*} \phi_2 + \phi_{2}^{*} \phi_1)^{n-1}
+...+D_{n+1} (|\phi_1|^2+|\phi_2|^2)^{n}\,.
\eea
It is easy to check that in these models also, the variable $X$
as defined by Eq. (\ref{xa}) is time independent, i.e. $\dot{X} = 0$.

\section{Straight-Gradient Superintegrable Quartic Schr\"odinger Dimer}

Finally, we  present a quartic straight-gradient superintegrable model. 
However, we must clarify that so far we have not been able to generalize
these results and obtain higher order superintegrable models based on
the straight-gradient Hamiltonian. Consider the Hamiltonian
\bea\label{z1}
&&H = i\gamma(|\phi_1|^2-|\phi_2|^2)+A_1(\phi_1 \phi_{2}^{*} 
+ \phi_2 \phi_{1}^{*})+A_2(|\phi_1|^2+|\phi_2|^2) \nonumber \\
&&+D_1 (|\phi_1|^2+|\phi_2|^2)^2  +D_2 |\phi_1|^2 |\phi_2|^2
+D_3(\phi_1 \phi_{2}^{*} + \phi_2 \phi_{1}^{*})^2\,.
\eea
This gives rise to the cubic dimer equations
\bea\label{z3}
&&i\frac{d\phi_1}{dt} = A_1 \phi_2 + A_2 \phi_1 + i\gamma \phi_1 
+2 D_1 (|\phi_1|^{2}+|\phi_2|^{2}) \phi_1 \nonumber \\
&&+D_2 |\phi_2|^2 \phi_1
+2 D_3 (\phi_{1}^{*} \phi_2 + \phi_{2}^{*} \phi_1)^{n-1} \phi_2\,,
\eea
\bea\label{z4}
&&i\frac{d\phi_2}{dt} = A_1 \phi_1 + A_2 \phi_2 - i\gamma \phi_1 
+2 D_1 (|\phi_1|^{2}+|\phi_2|^{2}) \phi_2 \nonumber \\
&&+D_2 |\phi_1|^2 \phi_2
+2 D_3 (\phi_{1}^{*} \phi_2 + \phi_{2}^{*} \phi_1)^{n-1} \phi_1\, . 
\eea

Using these field equations, it is easy to work out $\dot{X}, \dot{Y}$
as defined by Eqs. (\ref{xa}). We find
\be\label{z5}
\dot{X} = -D_2 YZ\,,~~~\dot{Y} = 2A_1 Z+(D_2+4D_3)XZ\,,~~~
\dot{R} = 2\gamma Z\,.
\ee
It is now easily checked that the two constants of motion are $C_1, C_2$
where
\be\label{z6}
C_1 = (aX-b)^2 + Y^2\,,~~~C_2 = R + f \sin^{-1}\left(\frac{aX-b}{\sqrt{C_1}}\right)\,,
\ee
and
\be\label{z7}
a = \sqrt{1+\frac{4D_3}{D_2}}\,,~~b = -\frac{2A_1}{aD_2}\,,~~
f = \frac{2\gamma}{aD_2}\,.
\ee
Summarizing, we then have a superintegrable dimer model based on the 
straight-gradient Hamiltonian with the three constants of motion being
$H, C_1, C_2$. 

\section{Conclusion} 

In recent years it has been widely realized that the PT-symmetric systems occupy
a position in between the dissipative and conservative systems \cite{ben}. In this 
connection, the PT-symmetric dimers have been playing an insightful role in
several areas \cite{kha}. It is thus of considerable importance to discover 
integrable and superintegrable dimers. Our work offers a step in that direction. 
We have presented a wide class of integrable oscillator type dimers whose 
Hamiltonian is of arbitrary even order. In addition, we presented a wide class 
of integrable and superintegrable nonlinear Schr\"odinger type dimers where 
again the Hamiltonian is of arbitrary even order.  

There are several interesting open problems. For example, can one construct
superintegrable oscillator type dimers? Secondly, can one consider superintegrable
Schr\"odinger type dimers of arbitrary even order based on straight-gradient 
Hamiltonians? Besides, it is useful to work out the detailed dynamics of these
models. Finally, it would be worthwhile knowing if some of these models could  
have physical relevance. We hope to address some of these issues in the near future. 

\section{Acknowledgments} 

We thank P. G. Kevrekidis and J. Cuevas-Maraver for discussions in the initial 
stages.  One of us (AK) is grateful to Indian National Science Academy (INSA) for 
the award of INSA senior Scientist position at Savitribai Phule Pune University. 
This work was supported in part by the U.S. Department of Energy.


\begin{thebibliography}{99}

\bibitem{ben} See for example, C. M. Bender, Rep. Prog. Phys. {\bf 70} 
(2007) 947 and references therein. 

\bibitem{opt} See, e.g., A. Ruschhaupt, F. Delgado and J. G. Muga, J. 
Phys. A: Math. Gen. {\bf 38} (2005) L171; K. G. Makris, R. El-Ganainy, D. N.
Christodoulides and Z. H. Musslimani, Phys. Rev. Lett. {\bf 100} (2008) 
10394; Int. J. Theor. Phys. {\bf 50} (2011) 1019; S. Klaiman, U. G\"unther
 and N. Moisejev, ibid {\bf 101} (2008) 080402; O. Bendix, R. Fleischmann, 
T. Kottos and B. Shapiro, ibid {\bf 103} (2009) 030402; S. Longhi, ibid 
{\bf 103} (2009) 123601; Phys. Rev. {\bf B 80} (2009) 235102; Phys. Rev.
{\bf A 81} (2010) 022102.

\bibitem{opex} A. Guo, G. J. Salamo, D. Duchesne, R. Morandotti, M. 
Volaiter-Ravat, V. Aimez, G. A. Siviloglou and D. N. Christodoulides, 
Phys. Rev. Lett. {\bf 103} (2009) 093902; C. E. R\"uter, K. G. Makris,
 R. El-Ganainy, D. N. Christodoulides, M. Segev and D. Kip, Nat. Phys.
 {\bf 6} (2010) 192; A. Regensburger, C. Bersch, M. A. Miri, G. 
Onishchukov, D. N. Christodoulides and U. Peschel, Nature {\bf 488} 
(2012) 167.

\bibitem{oscdim} C. M. Bender, B. Bernston, D. Parker and E. Samuel, 
Amer. J. Phys. {\bf 81} (2013) 173; B. Peng, S. K. \"Ozdemir, F. Lei, 
F. Monifi, M. Gianfreda, G. L. Long, S. Fan, F. Nori, C. M. Bender and
L. Yang, Nature Phys. {\bf 10} (2014) 394.

\bibitem{ele} J. Schindler, A. Li, M. C. Zheng, F. M. Ellis and T. Kottos,
 Phys. Rev. {\bf A 84} (2011) 040101; J. Schnidler, Z. Lin, J. M. Lee, 
H. Ramezani, F. M. Ellis and T. Kottos, J. Phys. A: Math. Theor. {\bf 45}
(2012) 444029.

\bibitem{cav} C.M. Bender, M. Gianfreda, S.K. \"Ozdemir, B. Peng and L. Yang, 
Phys. Rev. A {\bf 88} (2013) 062111.

\bibitem{kha} J. Cuevas, P.G. Kevrekidis, A. Saxena and A. Khare, Phys. Rev. A 
{\bf 88} (2013) 032108; I.V. Barashenkov and M. Gianfreda, J. Phys. A: Math. 
Theor. {\bf 47} (2014) 282001; J. Cuevas, A. Khare, P.G. Kevrekidis, H. Xu 
and A. Saxena, arXiv:1409.7218, Int. J. Theor. Phys. (2015), in press.
  
\bibitem{ba} I. V. Barashenkov, D. E. Pelinovsky and P. Dubard, J. Phys.
A: Math. Theor. {\bf 48} (2015) 325201.   

\bibitem{super} E. McSween and P. Winternitz, J. Math. Phys. {\bf 41}, 
2957 (2000).  



\end{thebibliography}
\end{document}